\documentclass{elsart}
\usepackage{graphicx}
\usepackage{lscape}

\begin{document}

\begin{frontmatter}

\title{\textit{dHybrid}: a massively parallel code for hybrid simulations of space plasmas}
%
%
\author[IST,RAL]{L. Gargat\'{e}\corauthref{lgar}}
\ead{luisgargate@cfp.ist.utl.pt}
\author[RAL,STRAT]{R. Bingham}
\author[ISCTE,IST]{R. A. Fonseca}
\author[IST]{L. O. Silva}
\corauth[lgar]{Corresponding author. Tel.: +351 21 8419336; fax: +351 21 8464455}
%
%
\address[IST]{GoLP/CFP, Instituto Superior T\'{e}cnico, Av. Rovisco Pais, Lisbon, Portugal}
\address[RAL]{Rutherford Appleton Laboratory, Chilton, Didcot, OXON, OX11 0QX, UK}
\address[STRAT]{Department of Physics, University of Strathclyde, Glasgow G4 ONG, Scotland}
\address[ISCTE]{DCTI, Instituto Superior de Ci\^{e}ncias do Trabalho e da Empresa, Av. For\c{c}as Armadas, Lisbon, Portugal}
%
%
\begin{abstract}
A massively parallel simulation code, called \textit{dHybrid}, has been developed to perform global scale studies of space plasma interactions. This code is based on an explicit hybrid model; the numerical stability and parallel scalability of the code are studied. A stabilization method for the explicit algorithm, for regions of near zero density, is proposed. 
Three-dimensional hybrid simulations of the interaction of the solar wind with unmagnetized artificial objects  are presented, with a focus on the expansion of a plasma cloud into the solar wind, which creates a diamagnetic cavity and drives the Interplanetary Magnetic Field out of the expansion region. The dynamics of this system can provide insights into other similar scenarios, such as the interaction of the solar wind with unmagnetized planets.
\end{abstract}
\begin{keyword}
hybrid codes \sep particle MHD codes \sep space plasmas \sep AMPTE \sep artificial atmospheres \sep solar wind
\PACS 52.65.Ww \sep 52.65.Kj \sep 96.50.Ek
\end{keyword}
\end{frontmatter}
%
%
\section{Introduction}
\indent 
To understand many space plasma scenarios, such as the solar wind interaction with cometary atmospheres or with unmagnetized planets (e.g. Mars) \cite{Dobe,Sagdeev} it is usually necessary to invoke the dynamics of ions. On one hand, MHD codes cannot always capture all the physics, (e.g. finite Larmor radius effects). On the other hand, full particle in cell (PIC) codes are computationally demanding and it is not always possible to simulate large scale space phenomena \cite{Dawson,birdsall,compsim}. Hybrid codes are useful in these problems, where the ion time scale needs to be properly resolved and the high frequency modes on the electron time/length scales do not play a significant role.

To address problems where the hybrid approximation is necessary we have developed the code \textit{dHybrid}, in which the ions are kinetic (PIC) and the electrons are assumed massless and treated as a fluid \cite{Lipatov,Brecht,Kazeminezhad}. There is also the possibility to use it as a particle MHD code \cite{Leboeuf,Kazeminezhad1}, by neglecting the ion kinetics. The present version of \textit{dHybrid} is fully parallelized, thus allowing for the use of massively parallel computers. Advanced visualization is performed by taking advantage of the close integration with the OSIRIS data analysis and visualization package \cite{RAF}. 

A stability analysis of the algorithm is performed, and stabilization mechanisms (associated with numeric instabilities due to very low density regions) with minimum impact on performance are discussed. The parallel scalability and performance of the algorithm is also presented.

The \textit{dHybrid} framework allows the study of a wide class of problems, including global studies of space plasma shock structures. This is illustrated by a set of simulations of an artificial gas release in the solar wind, depicting the AMPTE release experiments. The relevance of the example chosen is due to its resemblance to the solar wind interaction with planetary/cometary exospheres (e.g. Mars and Venus)\cite{Omidi,Bingham2}, thus illustrating the possible scenarios to be tackled with \textit{dHybrid}.

In the following section we describe the hybrid model used in \textit{dHybrid}. Its numerical implementation, focusing on stability and parallel scalability, is presented in section \ref{numerics}. We then describe the key features of the shock structures formed by the solar wind interaction with an unmagnetized artificial atmosphere. Comparison between present three dimensional simulations and previous 2D simulations \cite{Bingham2,Kazeminezhad2} are also presented in section \ref{applic}. Finally, we state the conclusions.
%
%
\section{Hybrid model and \textit{dHybrid}}
\label{theory}
\indent
Hybrid models are commonly used in many problems in plasma physics (for a review see, for instance, ref. \cite{Lipatov}). When deriving the hybrid set of equations the displacement current is neglected in Amp\`{e}re's Law and the kinetics of electrons is not considered. Various hybrid approximations can be considered if electron mass, resistivity and electron pressure are included or not in the model. Quasi-neutrality is also implicitly assumed. The appropriate approximation is chosen in accordance to which time scales and spatial scales are relevant for the dynamics of the system.

In \textit{dHybrid}, the electron mass, the resistivity and the electron pressure are not considered, but due to the code structure, such generalization is straightforward. Shock jump conditions (i.e. Rankine-Hugoniot relations) are altered by implicitly neglecting the electron temperature. The differences are significant when the $\beta$ of the plasma dominating the shock is high and in this case care should be taken when analyzing results. The electric field under these conditions is thus $\vec{E}=-\vec{V_e}\times\vec{B}$, in which $\vec{V_e}$ is the electron fluid velocity. The electric field is perpendicular to the local magnetic field, since the massless electrons short-circuit any parallel component of the electric field, and it can be determined from
\begin{equation}
\vec{E}=-\vec{V}\times\vec{B}+\frac{1}{n e \mu_{0}}\left(\nabla\times\vec{B}\right)\times\vec{B}
\label{eq:efield}
\end{equation}
where $\vec{V}=\frac{1}{n}\int f_i\,\vec{v}\, d\vec{v}$ is the ion fluid velocity. The magnetic field is advanced in time through Faraday's law where the electric field is calculated from eq. (\ref{eq:efield}). Ions in the hybrid model have their velocity determined by the usual Lorentz force. The Boris particle pusher is used, using the electric field and the magnetic field to advance velocities in time \cite{Boris}. In the particle MHD model one uses
\begin{equation}
\frac{\d\vec{v}}{\d t}=\frac{1}{\mu_0 n M}\left(\nabla\times\vec{B}\right)\times\vec{B}+\frac{k_{\mathrm{B}} T}{n M}\nabla n
\label{eq:MHDpushv}
\end{equation}
to determine individual particle velocities, where the second term on the right hand side is the pressure term for the ions, assuming an adiabatic equation of state, and where $k_{\mathrm{B}}$ is the Boltzmann constant. Ion fluid velocity is then obtained in the usual manner, integrating over the velocities \cite{Leboeuf}.

The ion species in \textit{dHybrid} are then always represented by finite sized particles to be pushed in a 3D simulation box. The fields and fluid quantities, such as the density $n$ and ion fluid velocity $\vec{V}$, are interpolated from the particles using quadratic splines \cite{Decyk} and defined on a 3D regular grid. These fields and fluid quantities are then interpolated back to push the ions using quadratic splines, in a self consistent manner. Equations are solved explicitly, based on a Boris pusher scheme to advance the particles \cite{Boris} in the hybrid approach, and a two step Lax-Wendroff scheme to advance the magnetic field \cite{birdsall,compsim}. Both schemes are second order accurate in space and time, and are time and space centered. 

The present version of \textit{dHybrid} uses the MPI framework as the foundation of the communication methods between processes, and the HDF5 framework as the basis of all diagnostics. The three-dimensional simulation space is divided across processes, and 1D, 2D and 3D domain decompositions are possible.

The code can simulate an arbitrary number of particle species and, for each of them, either the particle MHD or the hybrid model can be applied. Periodic boundary conditions are used for both the fields and the particles, and ion species are simulated with arbitrary charge to mass ratios, arbitrary initial thermal velocity and spatial configurations. This flexibility allows for simulations where only the kinetic aspects of one of the ion species is followed.

Normalized simulation units are considered for all the relevant quantities. Time is normalized to $\lambda_{\mathrm{pi}}/c_s$, space is  normalized to $\lambda_{\mathrm{pi}}$, mass is normalized to the proton mass $m_p$ and charge is normalized to the proton charge $e$, where $\lambda_{\mathrm{pi}}=c/\omega_{\mathrm{pi}}$ is the ion collisionless skin depth with $\omega_{\mathrm{pi}}=\sqrt{n_{0} e^{2}/\epsilon_{0}m_p}$ and where $c_s=\sqrt{(k_{\mathrm{B}} T_{i}+k_{\mathrm{B}} T_{e})/m_p }$ is ion the sound velocity. In this system the magnetic field is normalized to $m_p\,c_s/e\,\lambda_{\mathrm{pi}}$ and the electric field is normalized to $m_p\,c_s^2/e\,\lambda_{\mathrm{pi}}$. Henceforth all equations will be expressed in these normalized units.

Using the described implementation of the hybrid model, \textit{dHybrid} can model a wide range of problems, from unmagnetized to magnetized plasmas in different configurations. 
\section{Stability and scalability of \textit{dHybrid}}
\label{numerics}
\indent
The stability criterion on the time-step for all the algorithm is determined by the Lax-Wendroff method, as this is usually more stringent than the stability condition for the Boris algorithm due to the rapid increase in the Alfv\`{e}n velocity as the density goes to zero. The discretized equation (\ref{eq:efield}) is
\begin{equation}
\vec{E}^{n}_{i,j,k}=-\vec{V}^{n}_{i,j,k}\times\vec{B}^{n}_{i,j,k}+\frac{1}{n^{n}_{i,j,k}}\left(\nabla\times\vec{B}^{n}_{i+1/2,j+1/2,k+1/2}\right)\times\vec{B}^{n}_{i,j,k}
\label{eq:nefield}
\end{equation}
and the two step space-centered and time-centered Lax-Wendroff scheme to solve Faraday's law is 
\begin{equation}
\vec{B}^{n+1/2}_{i+1/2,j+1/2,k+1/2}=\vec{B}^{n}_{i+1/2,j+1/2,k+1/2}-\frac{\Delta\,t}{2}\left(\nabla\times\vec{E}^{n}_{i,j,k}\right)
\label{eq:nbfield1}
\end{equation}
\begin{equation}
\vec{B}^{n+1}_{i,j,k}=\vec{B}^{n}_{i,j,k}-\Delta\,t\left(\nabla\times\vec{E}^{n+1/2}_{i+1/2,j+1/2,k+1/2}\right)
\label{eq:nbfield2}
\end{equation}
where $\Delta t$ represents the time step, the $1/2$ index represent values displaced by half cell size, where $i$, $j$ and $k$ represent grid points along $x$, $y$ and $z$ and $n$ represents the iteration step. These equations thus require the use of staggered grids \cite{Yee}, where the displaced terms are calculated using an average of the eight neighbor values around a given point. 

The general layout of the \textit{dHybrid} algorithm is as follows. One starts of with particle positions at time $n$, velocities at time $n-1/2$ and $n$ (interpolated from time $n-1/2$), and the magnetic field at time $n$ grid 1 (position indexes $i,j,k$). In step (i) $V^n_{i,j,k}$, $n^n_{i,j,k}$ and $B^n_{i+1/2,j+1/2,k+1/2}$ are calculated from particles velocities, positions and from $B$ values in grid 1, (ii) electric field is calculated at $E^n_{i,j,k}$ from eq. (\ref{eq:nefield}), (iii) the magnetic field $B^{n+1/2}_{i+1/2,j+1/2,k+1/2}$ is calculated from eq. (\ref{eq:nbfield1}), (iv) particle velocites are calculated at $v^{n+1/2}$ using the Boris algorithm, positions at $x^{n+1}$ and $x^{n+1/2}$ are calculated from $x^{n+1}=x^n+\Delta t\, v^{n+1/2}$ and $x^{n+1/2}=(x^n+x^{n+1})/2$ and density and fluid velocity are calculated in grid 2: $n^{n+1/2}_{i+1/2,j+1/2,k+1/2}$ and $V^{n+1/2}_{i+1/2,j+1/2,k+1/2}$. In step (v) the magnetic field is calculated at $B^{n+1/2}_{i,j,k}$ from grid 2 values and then (vi) $E^{n+1/2}_{i+1/2,j+1/2,k+1/2}$ is calculated using eq. (\ref{eq:nefield}) displaced half grid cell, (vii) the magnetic field is advanced in time to $B^{n+1}_{i,j,k}$ using  eq. (\ref{eq:nbfield2}) and finally, (viii) particle velocities are advanced via Boris algorithm to $v^{n+1}$.

To obtain the Courant-Friedrichs-Levy stability condition, linearized versions of eq. (\ref{eq:nefield}) through eq. (\ref{eq:nbfield2}) are used considering constant density, arbitrary velocities and parallel propagating waves relative to the background magnetic field. The equations are then Fourier transformed to local grid modes, parallel plane waves $\sim\mathrm{e}^{\mathrm{i}\,\left(k_x\,\Delta x\,i-n\,\omega\,\Delta t\right)}$, where $\Delta x$ is the cell size in $x$.

An amplification matrix relating $\vec{B}^{n+1}_{i,j,k}$ with $\vec{B}^{n}_{i,j,k}$ is then obtained. Requiring that all the eigenvalues of the amplification matrix be $\lambda_i\leq 1$, yields the stability criterion
\begin{equation}
\Delta t\leq\frac{n\,\Delta x^2}{\left| n\,V_x\,\Delta x\pm B\right|}
\label{eq:courant}
\end{equation}
where $n$ is the background density, $B$ is the constant magnetic field, $V_x$ is the ion fluid velocity along $x$ and the two signs are due to the two different eigenvalues. The condition (\ref{eq:courant}) sets a limit on the time step or, inversely, given a time step a limit on the lowest allowable density which can be present in a grid cell. We stress that all quantities in eq. (\ref{eq:courant}) are expressed in normalized units. Using the same calculation method, a stability criterion was found in \cite{Kazeminezhad3} for similar field equations using a different implementation of the Lax-Wendroff algorithm. The stability criterion however is not the same since the specifics of the numerical approach differ: our implementation, described in \cite{Lipatov}, makes use of staggered grids to improve accuracy and guarantee that the equations are always space centered. 

As can be seen from eq. (\ref{eq:nefield}), under the MHD and hybrid models, the algorithm breaks in regions where the density is close to zero \cite{Hewett}. The problem is either physical, if particles are pushed out of a given region of space by a shock or other means, or it can be a numerical artifact due to poor particle statistics. One method to tackle this problem is by an implicit calculation of the electric field \cite{Lipatov}, which requires an iteration method to solve the electric field equation. One other method, discussed in \cite{Hewett,Arber} involves the use of an artificial resistivity.

If the problem is physical, it can be avoided by considering that if no particles exist in a given region of space, then both the charge density $\rho$ and current density $\vec{J}$ are zero, and the electric field is defined by $\nabla^{2}\vec{E}=0$. This equation has to be solved only on volumes where the density goes below a given stability threshold. This region can span several processes and thus will involve several communication steps. Fast elliptic solvers can be used to solve this equation, although the complex vacuum/plasma boundaries that arise complicate the problem.

Usually it is found that the problem is numerical in nature and it is only due to the limited number of computational particles per mesh cell used in the simulation. Thus, if in any given mesh cell there are too few particles, the density drops and the Alfv\`{e}n velocity $v_A=B/\sqrt{\mu_0 n m}$ increases, breaking the stability criterion for the field solver. 

Three methods are considered to maintain the stability of the algorithm: (i) the number of particles per cell can be increased throughout the simulation, (ii) the time step can be decreased, and (iii) a non-physical background density can be added as needed in the unstable zones, rendering them stable. The two former options are obvious, and yield meaningful physical results at expense of computational time. The last solution is non-physical and, thus, can yield non-physical results. In short, the first two options are chosen and the last one is implemented to be used only as last resort. Each time the electric field is to be calculated, the density in each mesh cell is automatically checked to determine if it is below the stability threshold, using eq. (\ref{eq:courant}), and is set to the minimum value if it is. The minimum density value is thus calculated using the time step, the grid size, and the local values for the magnetic field and fluid velocity, minimizing the impact of using a non-physical solution. Testing showed that as long as the number of cells that are treated with this method is kept low enough (up to $0.1\%$ of the total number of cells), the results do not change significantly.   

The approach followed here guarantees good parallel scalability of the algorithm since the algorithm is local. The overall algorithm was also designed as to be as local as possible and to minimize the number of communication steps between processes. This was accomplished by joining several vectors to transmit at the same step and it resulted in the following parallel scheme: (i) after the first step in the main algorithm guard cells of $n^{n}_{i,j,k}$, $V^{n}_{i,j,k}$ and $B^{n}_{i+1/2,j+1/2,k+1/2}$ are exchanged between neighboring processes, (ii) guard cells of $E^{n}_{i,j,k}$ are exchanged, (iii) guard cells of $B^{n+1/2}_{i+1/2,j+1/2,k+1/2}$ are exchanged, (iv) particles that crossed spacial boundaries are exchanged between neighboring processes, (v) guard cells of $n^{n+1/2}_{i+1/2,j+1/2,k+1/2}$, $V^{n+1/2}_{i+1/2,j+1/2,k+1/2}$ and $B^{n+1/2}_{i,j,k}$ are exchanged, (vi) guard cells of $E^{n+1/2}_{i+1/2,j+1/2,k+1/2}$ are exchanged and finally (vii) guard cells of $B^{n+1}_{i,j,k}$ are exchanged.

Scalability of \textit{dHybrid} was studied on a Macintosh dual G5 cluster at $2\,\mathrm{GHz}$, interconnected with a Gigabit ethernet network. The particle push time is $1.7\,\mathrm{\mu s}$ per particle, and the field solver time is $8.4\%$ of the total iteration time for 1000 iterations on a single process. 

One plasma species is initialized evenly across the simulation box with a drift velocity of $5\,\mathrm{c_s}$, a thermal temperature of $0.1\,\mathrm{c_s}$ and a charge to mass ratio of $1$ (protons). A perpendicular magnetic field with constant intensity of $B_0=5\,\mathrm{m_p\,c_s\,e^{-1}\,\lambda_{pi}^{-1}}$ is set across the box. The benchmark set up consists of two different "parallel" scenarios.
\begin{figure}[ht]
\centering 
\includegraphics{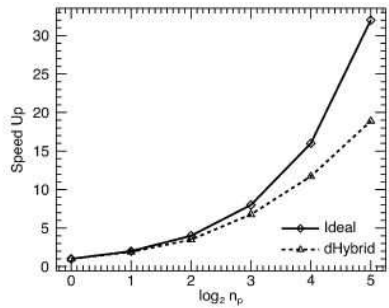}
\label{fig:speedup1}
\caption{Code speed up when problem size is kept fixed for all simulations (independent of $\mathrm{n_p}$).}
\end{figure}

In the first scenario a 96 cell cubic grid is used, with 8 particles per cell, all diagnostics off, and 1000 iterations are performed. The simulation space is then evenly split up among the number of processes in each run, in a 1D partition. The average time per iteration is taken, relative to the time per iteration in a single processor. Fig. 1 compares the ideal speed up against the achieved results. The minimum speed up obtained is $\sim60\%$ when using 32 processors. We observe that in this case the maximum problem size that could be set on one machine is limited by memory considerations and therefore, when the problem is split up by 32 processors, the problem size per processor is much smaller and the communication time relative to the overall iteration time starts to penalize the code performance, reaching $40\%$ of the loop time for 32 processors. 
\begin{figure}[ht]
\centering 
\includegraphics{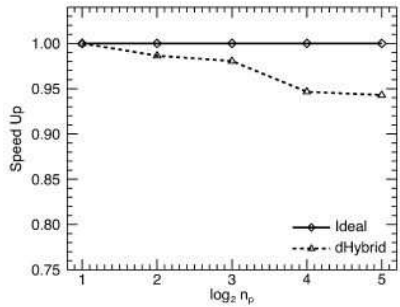}
\label{fig:speedup2}
\caption{Code speed up when problem size is doubled when doubling the number of processes.}
\end{figure}

In the second scenario, the problem size increases proportionally to the number of processors used. Fig. 2 shows the results for runs with 2, 4, 8, 16 and 32 processor runs. The efficiency in this case is $94\%$ showing good parallel scaling, as expected, with similar efficiency as other state-of-the-art massively parallel codes \cite{RAF,RAF2}. The total communication time takes $8.5\%$ of the total iteration time in this case, thus showing that this problem is more indicative of the parallel efficiency of the algorithm; the penalty for scaling up to many processors is not significant. Other test runs with the same setup but with no magnetic field were considered, and the efficiency in this case was higher with a constant value of $\sim99\%$ for all the runs. This indicates that the drop in efficiency patent in Fig. 2 is mainly due to particle load balancing across processes, induced by the magnetic field, which makes particles have a Larmor radius of $\sim10\%$ the simulation box size in the x dimension.

%
%
\section{3D simulations of unmagnetized objects}
\label{applic}
\indent
As a test problem for \textit{dHybrid}, we have modeled the interaction of the solar wind with an unmagnetized object, mimicking the AMPTE release experiments, thus allowing the validation of the code against the AMPTE experimental results and other codes \cite{Kazeminezhad3,Valenzuela,Haerendel,chapman,Harold,Delamere}.

The AMPTE experiments consisted of several gas (Lithium and Barium) releases in the upstream solar wind by a spacecraft orbiting the earth \cite{Bernhardt,Hassan,Bingham1}. After the release, the expanding cloud of atoms is photoionized by the solar ultraviolet radiation, thus producing an expanding plasma and forming an obstacle to the flowing solar wind. The solar wind drags the Sun's magnetic field corresponding, at $1\,\mathrm{AU}$, to a fairly uniform background field with a magnitude of about $10\,\mathrm{nT}$ in the AMPTE experiments. The measured solar wind density was $n_{sw}=5\,\mathrm{cm^{-3}}$, flowing with a velocity of $v_{sw}\sim540\,\mathrm{km/s}$ and having an ion acoustic speed of $c_s\sim53\,\mathrm{km/s}$. The cloud expansion velocity was $\sim 2.1\,\mathrm{km/s}$.

A number of codes were developed to study the AMPTE release experiments in the solar wind, most of them 2D codes. These simulations show that the MHD model lacked certain key physics \cite{conference,Bollens}. The correct modeling of the cloud dynamics can only be obtained in hybrid simulations, because in the AMPTE releases, the ion Larmor radius of the solar wind particles is in the same order of magnitude of the cloud size itself. The problem is intrinsically kinetic and can only be fully assessed in a 3D simulation as this yields realistic field decays over space providing realistic dynamics for the ions. This is also of paramount importance for the ion pick up processes in planetary exospheres \cite{Luhmann,Sauer}.

In our simulations, the background magnetic field was set to $B_0=0.773\,\mathrm{nT}$  with a solar wind velocity of $v_{sw}=80.4\,\mathrm{km/s}$ and with a cloud expansion velocity of $v_c=16.1\,\mathrm{km/s}$. The relative pressure of the solar wind plasma due to the embedded magnetic field (low $\beta$ plasma), and the relative pressure of the plasma cloud due to the expanding velocity of the cloud (high $\beta$ plasma), were kept fixed. These two pressures control the shock structure, and determine the physical behavior of the system. 

The simulations were performed on a computational cubic grid of $300^3$ cells, 12 solar wind particles per cell, a time step of $t=6.25\times 10^{-3}\,\mathrm{\lambda_{pi}/c_s}$ and a cubic box size of $150\,\mathrm{\lambda_{pi}}$ in each dimension. A 2D parallel domain decomposition in x and y was used. The normalizing quantities are $\lambda_{pi}=102\,\mathrm{km}$ for spatial dimensions, $c_s=53.6\,\mathrm{km/s}$ for the velocities, $\lambda_{pi}/c_s=1.9\,\mathrm{s}$ for the time, $5.49\,\mathrm{nT}$ for the magnetic field and $n_0=5\times10^6\,\mathrm{m^{-3}}$ for the density. In the simulations, the solar wind is flowing from the $-x$ side to the $+x$ side and the magnetic field is perpendicular to this flow, along the $+z$ direction. As the cloud expands, a magnetic field compression zone is formed in the front of the cloud. The solar wind ions are deflected around the cloud due to the magnetic barrier, drift in the $-y$ direction piling up in the lower region of the cloud, and are accelerated in this process. 
\begin{figure}[ht]
\centering 
\includegraphics{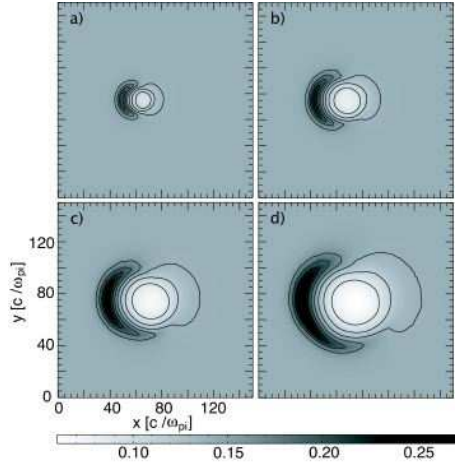}
\label{fig:bslice}
\caption{Slice of the magnetic field magnitude at $z=75\,\mathrm{\lambda_{pi}}$, for times a) $t=6.25\,\mathrm{\lambda_{pi}/c_s}$, b) $t=12.5\,\mathrm{\lambda_{pi}/c_s}$, c) $t=18.75\,\mathrm{\lambda_{pi}/c_s}$ and d) $t=25\,\mathrm{\lambda_{pi}/c_s}$. Iso-lines of the magnetic field magnitude are also shown.}
\end{figure}

The magnetic field assumes a particular importance to test the model as it has a very clear signature, characteristic of the AMPTE experiments. In Fig. 3 the magnetic field evolution is shown in the center plane of the simulation. As the plasma bubble expands a diamagnetic cavity is formed due to the outward flow of ions that creates a diamagnetic current. A magnetic field compression zone is also evident - the kinetic pressure of the cloud ions drives the Interplanetary Magnetic Field outwards, creating the compressed magnetic field. These results reproduce 2D run results obtained in previous works \cite{Bingham2,Kazeminezhad2}, and are in excellent agreement with the AMPTE experiments. 
\begin{figure}[ht]
\centering 
\includegraphics{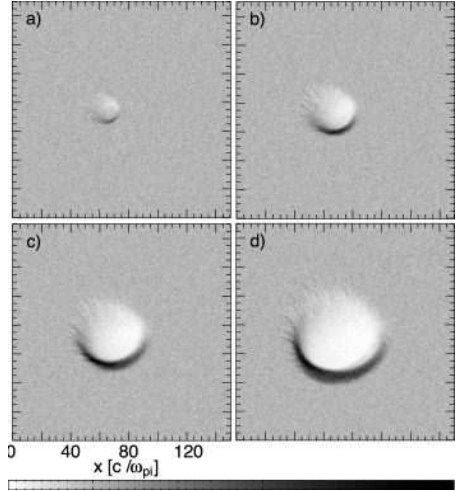}
\label{fig:swslice}
\caption{Slice of the solar wind density at $z=75\,\mathrm{\lambda_{pi}}$, for times a) $t=6.25\,\mathrm{\lambda_{pi}/c_s}$, b) $t=12.5\,\mathrm{\lambda_{pi}/c_s}$, c) $t=18.75\,\mathrm{\lambda_{pi}/c_s}$ and d) $t=25\,\mathrm{\lambda_{pi}/c_s}$. A pile up zone is visible along with a density depleted zone.}
\end{figure}

In Fig. 4, taken at the same  time steps, it is visible that the solar wind ions are being pushed out of the cloud area. The solar wind coming from the $-x$ direction is deflected around the magnetic field pile up zone and drifts in the $-y$ direction. This is due to the the electric field generated inside the cloud, dominated by the outflowing ions, that creates a counter clock-wise electric field. This electric field, depicted in Fig. 5, is responsible for the solar wind ion drift in the $-y$ direction. The same electric field also affects the cloud ions, that are pushed out in the $+y$ side of the cloud, and pushed back in on the other side. The ejection of ions in the $+y$ side, known as the rocket effect \cite{Haerendel}, is one of the reasons of the reported bulk cloud drift in the $-y$ direction due to momentum conservation and is thoroughly examined along with other effects in \cite{Kazeminezhad2}.

One other interesting aspect is that as the simulation evolves, there are regions of space in which the density drops, making this test problem a good choice to test the low density stability problem resolution. It was found that density dropped below the stability limit not in the center of the cloud, but behind it, in the downwind area in the $+x$ side of the cloud (Fig. 4). In the center of the cloud although the solar wind is pushed out, high density is observed, due to the presence of the cloud ions. It was also found that this was an example of a low density due only to poor particle statistics, as it happened only when 8 particles per cell were used and was eliminated by increasing the number of particles per cell to 12. The results, however, were very similar in the two runs with 8 particles per cell versus the 12 particle per cell run, due to the non-physical stabilization algorithm used.

\begin{figure}[ht]
\centering 
\includegraphics{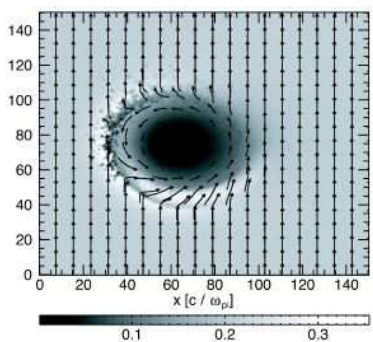}
\label{fig:efield}
\caption{Slice of the electric field at $z=75\,\mathrm{\lambda_{pi}}$, for time $t=25\,\mathrm{\lambda_{pi}/c_s}$. Field vectors show counter clockwise rotating electric field.}
\end{figure}

The total ion fluid velocity is shown in Fig. 6. The dark isosurface in the outer part of the cloud represents fluid velocities in the order of $2\,\mathrm{c_s}$. This is a clear indication of an acceleration mechanism acting on these particles, which is due to the same electric field generated by the outflowing cloud ions. 
\begin{figure}[ht]
\centering 
\includegraphics{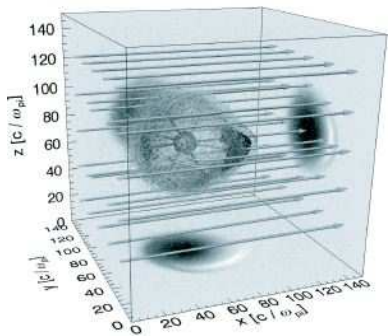}
\label{fig:vfld}
\caption{Ion fluid velocity field lines, iso-surfaces and projections. The darker iso-surface represents fluid velocities of $\sim2\,\mathrm{c_s}$.}
\end{figure}

The observations made at the time of the AMPTE releases match the results from the simulations.
A shock like structure, as in the simulations, was observed and reported along with a large diamagnetic cavity, coincident with the peak cloud density area \cite{Kazeminezhad3,Bingham1}. Both the instability in the cloud expansion and the rocket effect responsible for the cloud recoil in the $\vec{v}\times\vec{B}$ direction, observed in the actual experiments and reported in several papers \cite{Bingham2,Kazeminezhad2,Delamere,Bernhardt}, were also observed in \textit{dHybrid} simulations.

Other features like ion acceleration on the $-y$ downwind side of the cloud, and the charge density pile up on the $-y$ side of the cloud, which were unobserved in previous simulations, were captured in our simulations due to the use of much higher resolutions. These effects are due to the cloud expansion that creates an electric field capable of deflecting the solar wind around the cloud and accelerate the solar wind particles.
%
%
\section{Conclusions}
\label{conc}
\indent
In this paper we have presented the code \textit{dHybrid}, a three dimensional massively parallel numerical implementation of the hybrid model. The stability of the algorithm has been discussed and a stabilization criterion with no impact on parallel scalability has been proposed and tested. The AMPTE release experiments were modeled, and the main physical features were recovered with \textit{dHybrid}. 

Zero densities on hybrid and MHD models are a source of instabilities. In the hybrid case these are usually numerical artifacts due to poor particle statistics. Numerical stability analysis of \textit{dHybrid} has been carried out and a constraint both on the time step and on the density minimum has been found. This constraint helps to effectively determine at run time where instabilities will occur and suppress them. 

The parallel scalability of the algorithm has been studied yielding a $94\%$ parallel efficiency for scaled problem sizes for typical runs with 32 processes, showing excellent scalability, an indication that the parallelization method is efficient in tackling larger problems. The zero density stabilized algorithm does not suffer from parallel performance degradation, thus avoiding the pitfalls of other solvers that require inter-process communication steps.

\textit{dHybrid} has been tested through comparison both with previous two dimensional codes and the experimental results from the AMPTE release experiments. The key features of the AMPTE release experiments are recovered by \textit{dHybrid}.

Parallel \textit{dHybrid} allows the full scale study of the solar wind interaction with unmagnetized objects. Similar problems, such as planetary exosphere erosion in Mars and in Venus can be tackled, and will be the topic of future papers. 
%
%
\begin{ack}
\indent
The authors wish to acknowledge Dr. J. N. Leboeuf for providing the code \textit{dComet} which led to the development of \textit{dHybrid}.
This work is partially supported by FCT (Portugal), the European Research Training Network on Turbulent Layers under European Commission contract HPRN-CT-2001-00314 and CCLRC Center for Fundamental Physics (UK).
\end{ack}
\end{document}